\documentclass[conference]{IEEEtran}
\IEEEoverridecommandlockouts
\usepackage{cite}
\usepackage{amsmath,amssymb,amsfonts}
\usepackage{algorithmic}
\usepackage{graphicx}
\usepackage{textcomp}
\usepackage{xcolor}
\def\BibTeX{{\rm B\kern-.05em{\sc i\kern-.025em b}\kern-.08em
    T\kern-.1667em\lower.7ex\hbox{E}\kern-.125emX}}
    
\usepackage{lipsum}

\usepackage{caption}

\usepackage{amsmath,graphicx}
\newlength{\fslength}
\makeatletter
\newcommand{\funnyP}{%
    \setlength{\fslength}{\f@size pt}%
    \reflectbox{${\mbox{P}}$}\hspace*{-.300\fslength}\mbox{${\mbox{P}}$}%
}
\newcommand{\funnyPbar}{%
    \setlength{\fslength}{\f@size pt}%
    \reflectbox{$\bar{\mbox{P}}$}\hspace*{-.300\fslength}\mbox{$\bar{\mbox{P}}$}%
}
\newcommand{\funnyPbarM}{%
    \setlength{\fslength}{\f@size pt}%
    \reflectbox{$\bar{\mbox{P}}$}\hspace*{-.300\fslength}\mbox{$\bar{\mbox{P}}_{M}$}%
}
\makeatother

\usepackage{stackengine}
\usepackage{array}

\begin{document}

\title{Toward Open Repository of Performance Portability of Applications, Benchmarks and Models\\
}

\author{\IEEEauthorblockN{Ami Marowka}
\text{Parallel Research Labs}\\
Petach-Tikva, Israel \\
amimar2@yahoo.com
}

\maketitle

\begin{abstract}
The adoption of heterogeneous computing systems based on diverse architectures to achieve exascale computing power has worsened the performance portability problem of scientific applications that were designed to run on these platforms.

\smallskip
To cope with the challenges posed by supercomputing, new performance portability frameworks have been developed alongside advanced methods and metrics to evaluate the performance portability of heterogeneous applications. However, many studies have shown that the new methods and metrics do not produce coherent results which yield clear conclusions that are required for designing the hardware and software architectures of tomorrow's supercomputing systems.

\smallskip
We outline a proposal to establish an open repository of performance portability of applications, benchmarks and models which will be standardized, objective, and based on strict operating and reporting guidelines. Such guidelines will ensure a fair, comparable and meaningful measure of the performance portability while the requirement for a detailed disclosure of the obtained results and the configuration settings will ensure the reproducibility of the reported results.
\end{abstract}

\smallskip
\begin{IEEEkeywords}
Performance Portability, Performance Efficiency, Metrics, SPEC
\end{IEEEkeywords}

\section{Introduction}
Emerging performance portability frameworks such as Kokkos \cite{Kokkos}, Raja \cite{RAJA} and SYCL \cite{SYCL} alongside mature heterogeneous high-level programming models such as OpenMP \cite{OpenMP}, OpenACC \cite{OpenACC} and MPI \cite{MPI} are the main software development infrastructures that will be available for software engineers to build scientific applications in the era of exascale computing.

The interplay between the never-ending demand for high performance applications, on the one hand, and the demand for portability and productivity of those applications, on the other hand, becomes more complex as hardware architectures become more heterogeneous. The performance portability frameworks developed in recent years haved shown impressive progress in everything related to functional portability with the appearance of high-level cross-platform programming models based on the approach of backend compilers, such as Kokkos, and a single-source C++ standard for heterogeneous computing, such as SYCL.

Despite all this impressive progress, performance portability still poses challenging technological issues to software and hardware architects. Dealing with these issues requires, first and foremost, an agreed definition for the term performance portability and agreed metrics for measuring and evaluating the degree of performance portability of heterogeneous applications, benchmarks, and higher-level heterogeneous programming models.

Furthermore, in order to measure performance portability in a way that it will be possible to compare different implementations of the same application in a meaningful and objective manner, clear and agreed upon guidelines and rules are needed for how the measurements should be performed and reported so that they can be reproduced. In addition, the results should be available and accessible to the High Performance Computing (HPC)  community in an open repository.

Of all the necessary requirements for having a methodological framework for measuring and comparing performance portability, it seems that regarding the definition of the term performance portability there is a broad consensus \cite{Pennycook2}:

\begin{quote} 
{\bf Definition: performance portability}

{\it A measurement of an application's performance efficiency for a given problem that can be executed correctly on all platforms in a given set.}
\end{quote}

The definition explicitly states that performance efficiency is the ultimate measure of performance portability.
Therefore, several approaches were proposed to measure performance efficiency alongside several metrics to calculate performance portability \cite{Pennycook2,Marowka2,Pennycook3,Marowka3,Marowka5}. And if we add to these facts that there is no agreed framework of rules and guidelines on how to measure and calculate performance efficiency and performance portability, then it is not surprising that it is not possible to draw informed insights from the dozens of studies that have been done in recent years, and it would not be an overstatement to claim that the current situation is a complete mess that can be reorganized.

This paper is intended to delineate a way to organize future studies of performance portability under an uniform framework of rules and guidelines for measuring, calculating and reporting performance portability of applications, benchmarks and performance portability frameworks. We demonstrate our approach using the Standard Performance Evaluation Corporation (SPEC) benchmarks \cite{SPEC} as a way to solve the disorganization that exists in this important research area. However, other similar frameworks can be appropriate alternative infrastructures for the ideas presented in this paper.

The main contribution of this paper lies in the novel idea of how to integrate the future studies of performance portability in an existing and dynamic framework that has proven itself over three decades and, as we will see later, it already has the basic definitions.
%that allow the ideas presented in this paper to be partially demonstrated.  
Furthermore, we would like to emphasize that in this paper we are only sketching the proposed framework and the examples we use to demonstrate the calculation of the performance portability are based only on the measurements that appear within the current SPEC repository. 
We would like to remind the reader that SPEC was designed to be a performance benchmarking framework for HPC platforms and not performance portability benchmarking framework. 
%and therefore we are currently limited from presenting all our ideas presented in this paper.

\smallskip
With that goal, we make the following contributions:
\begin{itemize}
\item{We present the main problems which cause the inconsistent measurement, calculation, and reporting of performance portability results in the studies that have been carried out in recent years and which yield inconsistencies.}
\item{We introduce new types of performance efficiencies in addition to the existing ones in order to enable analysis of performance portability of applications and models from different perspectives.}
\item{We demonstrate the calculation of performance portability of applications and benchmarks based on currently published SPEC performance measurements.}
\end{itemize}

\smallskip
The rest of the paper is structured as follows. Section 2 presents the motivation to establish an orderly framework for examining the performance portability of applications. Section 3 presents related studies. Section 4 presents the SPEC benchmarks framework. Section 5 presents our suggestion to integrate in SPEC the evaluation of the performance portability. Section 6 demonstrates the calculation of performance portability
of applications that are currently appear in SPEC and Section 7 presents conclusions.

\section{Motivation}

In this section, we present the current main performance portability issues that call for organizing this research field in order to enable informed conclusions to be drawn from future studies. Furthermore, due to these issues there is fundamental motivation to maintain a rigid framework of rules and regulated measurement mechanisms for future studies of performance portability whose results will be stored in an open repository accessible to the HPC community.

The report of the first Department of Energy (DOE) Performance, Portability and Productivity annual meeting  in 2016 showed clearly that there is no consensus on a workable definition of the performance portability term \cite{DOE}. This situation led researchers to propose the definition presented in the introduction and which has been widely accepted in the HPC community. This meeting motivated Pennycook et. al. to propose  the {{\funnyP}} metric to calculate the performance portability based on the harmonic mean \cite {Pennycook2}. But this metric proved itself to be problematic as was articulated in many studies
\cite{Marowka2,Marowka5,Dreuning,Daniel,Bertoni}.

The main claims against the {\funnyP} metric were that it is unintuitive, unfamiliar, loses information, difficult to use, and the performance portability scores it yields are unrealistic. Therefore, the {\funnyPbar} metric based on the arithmetic mean was proposed, which actually solved the above problems and yielded much more realistic results without losing information \cite{Marowka2}. The designers of the {\funnyP} metric accepted some of the claims but left the rest of the problems unaddressed  \cite{Pennycook3}. Currently, the situation is that there are studies that still use the {\funnyP} metric but not according to the original definition in order to avoid the aforementioned problems \cite{Deakin3}. For example, based on the {\funnyP} metric, if one platform does not support an application, it suggests that the performance portability of the application is zero. This, however, just does not make sense because there is always a platform out there that does not support a given application. Therefore, what actually happens is that the metric is used in such a way that only those platforms which support the application are taken into account \cite{Deakin3,Bertoni}. Otherwise, the performance portability scores will be zero and thus meaningless, as has happened in many studies \cite{Hsu,Harrell,Deakin}. The current situation is that there are currently two metrics, including one that is still controversial, which is an undesirable situation.  

Another issue is related to the performance efficiency approaches that are currently in use: application efficiency and architectural efficiency approaches. The widespread claim is that it is not clear which approach to use, since each one produces different results \cite{Dreuning}. Although the two approaches complement each other, the situation is still far from clear for many researchers . The best indication for this claim is that, to the best of our knowledge, there has not yet been even a single study that has used both approaches for a given application-platform pair and then performed an appropriate analysis of the results. In Section 5 we present different types of each of the approaches that can be included in the SPEC framework, but not necessarily all of them will be mandatory. Undoubtedly, a combination of types from both approaches provides more insights of the performance portability of a given application.

\section{Related Work}

This section presents a few related studies that have criticized the {\funnyP}  metric and those that have proposed solutions for improving the metric.
Furthermore, this section elaborates on the issues presented in the previous section by presenting the misunderstandings of different researchers regarding how to calculate the performance portability of applications. 

Dreuning et al. convincingly presented some of the dilemmas that the  {\funnyP} metric poses to developers and the ambiguity of the results obtained \cite {Dreuning}. They demonstrated the usability and the usefulness of the {\funnyP} metric by implementing five OpenACC applications using a set of three platforms (one CPU and two GPUs). 
The first question they asked themselves was: Which measure to use, bandwidth or operational throughput? The solution they found was to use the Roofline model \cite{Roofline} to calculate the ratio of the application and the hardware operational intensity values to determine whether the application was compute- or memory-bound and accordingly whether to use bandwidth or operational throughput.
The second question was: Which performance efficiency to use, application or architectural efficiency? From analyzing the results of their experiments, they concluded that to assess whether the performance of a given application can be improved further, architectural efficiency alone is not sufficient, and a diagnosis of what the application efficiency provides is also required.

They also noted that the harmonic mean tracks the low values of the CPUs even though the values of the GPUs are significantly higher. 
We showed that this observation is typical of the {\funnyP} metric, but not of {\funnyPbar}, which is why we recommend always to present the scores for CPUs and GPUs separately \cite{Marowka2}.

Siklosi et al. examined the performance of Stencil applications on hybrid CPU-GPU systems \cite {Siklosi}. They found that using the {\funnyP} metric to calculate the performance portability of applications is not intuitive. In their opinion, the reason for this is that if architectural efficiency is used, then the {\funnyP} metric tends to track the low values and therefore the improvement of a hybrid system is not reflected in the calculated {\funnyP} score. However, when using application efficiency, a hand-tuned baseline implementation is required, which to the best of their knowledge does not exist.

Daniel and Panetta showed that the {\funnyP} metric is easily affected by the problem size \cite {Daniel}. To address this susceptibility, they proposed an alternative metric called {\it Performance Portability Divergence ($P_{D}$)} as the arithmetic mean of RMS divergences across a set of platforms $H$:

%\smallskip
%\begin{align}
\begin{center}
$P_{D} = \frac{\sum_{i \in H} {\Delta_{RMS}}}{|H|}$       
\end{center}
%\end{align} 
%\smallskip

where the divergence RMS, ${\Delta_{RMS}}$, is the root mean square of performance distances between a set of input sizes and {\it Performance Distance} is the relative error in performance measure between two applications solving the same problem with the same platform and input size. The performance measure used by Daniel and Panetta  is the application efficiency.

The $P_{D}$ metric is different from the {\funnyP} and {\funnyPbar} metrics. 
It does not capture the performance and portability of an application across platforms. The {\funnyP} and {\funnyPbar} metrics calculate the average performance efficiencies of a given application on top of a given architecture set. On the other hand, the $P_{D}$ metric calculates the average variability of the performance efficiencies of a number of input sizes of a given application on top of a given set of architectures. These are therefore two distinct products. The $P_{D}$ metric can be a complementary metric to {\funnyP} and {\funnyPbar} that shows the variance obtained from different input sizes.

Sedova et al. proposed a performance portability metric denoted by the symbol $PP_{MD}$, where $MD$ stands for Molecular Dynamics \cite {Sedova}. It measures the contributions of non-portable components to an application's performance. The $PP_{MD}$ metric is the harmonic mean of the speedups of the application's components that are low-level, optimized and non-portable.   

Sedova et al. do not explain why they chose to use the harmonic mean. Unlike the {\funnyP} and {\funnyPbar} metrics, the $PP_{MD}$ metric is calculated for a particular architecture rather than a set of architectures. The $PP_{MD}$ metric purports to evaluate performance portability but in practice it measures the price in performance that must be paid to make the application portable.

Bertoni et al. studied how several OpenCL implementations of the Rodinia Benchmarks performed across three platforms and used the {\funnyP} metric to estimate the performance portability of the tested implementations \cite {Bertoni}.
They claimed that the {\funnyP} metric was insufficient for this purpose because it scored different implementations equally despite the fact that their performance efficiencies were very different. Therefore, they proposed to measure the standard deviation of the performance efficiencies to add another perspective on the performance efficiencies distribution across platforms.

It is argued here that using the {\funnyPbar} metric improves the diagnoses. Table \ref{table:t1} shows the performance efficiencies of the various implementations on the platforms used and the scores of the {\funnyP} and {\funnyPbar} metrics side by side along with their standard deviations. Clearly, the scores of the {\funnyPbar} metric differentiate better which of the implementations have better performance portability and it also more reliably reflects the performance efficiencies from which the 
{\funnyPbar} values are derived. Pay particular attention to how the scores of the SC and HS kernels have changed significantly. 

%---

\begin{table}
\caption{Comparison between the performance portability scores obtained by {\funnyP} vs. {\funnyPbar} metrics in the study from \cite{Bertoni}.}
\begin{center}
%\footnotesize
%\resizebox{\columnwidth}{0.08\textheight}{
\resizebox{\columnwidth}{!}{% 
\setlength\extrarowheight{2pt}
\begin{tabular}{|l|l|l|l|l|l|l|l|}

\hline  
\multicolumn{1}{|c|}{} 
&\multicolumn{3}{c|}{Platforms} 
&\multicolumn{4}{c|}{}  \\ \cline{1-8}

\multicolumn{1}{|c|}{Kernel}
&\multicolumn{1}{c|}{SKX}
&\multicolumn{1}{c|}{Gen9}
&\multicolumn{1}{c|}{V100}
&\multicolumn{1}{c|}{\addstackgap[3pt]{\funnyP}} 
&\multicolumn{1}{c|}{S.D.(HM)} 
&\multicolumn{1}{c|}{{\funnyPbar}}
&\multicolumn{1}{c|}{S.D.(AM)}  \\ \cline{1-8}

\multicolumn{1}{|c|}{LUD}
&\multicolumn{1}{c|}{35.89\%}
&\multicolumn{1}{c|}{48.71\%}
&\multicolumn{1}{c|}{49.80\%}
&\multicolumn{1}{c|}{43.81\%} 
&\multicolumn{1}{c|}{8.39} 
&\multicolumn{1}{c|}{44.80\%}
&\multicolumn{1}{c|}{6.31}  \\ \cline{1-8}

\multicolumn{1}{|c|}{BP-AW}
&\multicolumn{1}{c|}{--}
&\multicolumn{1}{c|}{81.73\%}
&\multicolumn{1}{c|}{91.66\%}
&\multicolumn{1}{c|}{86.41\%} 
&\multicolumn{1}{c|}{7.00} 
&\multicolumn{1}{c|}{86.67\%}
&\multicolumn{1}{c|}{4.96}  \\ \cline{1-8} 

\multicolumn{1}{|c|}{SC}
&\multicolumn{1}{c|}{9.97\%}
&\multicolumn{1}{c|}{50.58\%}
&\multicolumn{1}{c|}{92.90\%}
&\multicolumn{1}{c|}{22.94\%} 
&\multicolumn{1}{c|}{25.93} 
&\multicolumn{1}{c|}{51.15\%}
&\multicolumn{1}{c|}{33.85}  \\ \cline{1-8}

\multicolumn{1}{|c|}{KNN}
&\multicolumn{1}{c|}{78.15\%}
&\multicolumn{1}{c|}{40.32\%}
&\multicolumn{1}{c|}{35.50\%}
&\multicolumn{1}{c|}{45.61\%} 
&\multicolumn{1}{c|}{16.82} 
&\multicolumn{1}{c|}{51.32\%}
&\multicolumn{1}{c|}{19.07}  \\ \cline{1-8}

\multicolumn{1}{|c|}{HS}
&\multicolumn{1}{c|}{16.47\%}
&\multicolumn{1}{c|}{96.03\%}
&\multicolumn{1}{c|}{72.40\%}
&\multicolumn{1}{c|}{35.33\%} 
&\multicolumn{1}{c|}{35.10} 
&\multicolumn{1}{c|}{61.63\%}
&\multicolumn{1}{c|}{33.36}  \\ \cline{1-8}

\end{tabular}
}
\end{center}
\label{table:t1}
\end{table}

Bertoni et al. chose to calculate the performance efficiencies in relation to the Roofline peak performance. They describe in detail the methodology used to construct the Roofline graphs, thus demonstrating how complex and exhausting the process is.

\section{SPEC Benchmarks}

In this section we present the SPEC benchmark suites that are relevant to the topic of the present paper. We will focus on describing the main set of run-rules to which an implementer needs to adhere when using these benchmarks for measuring the performance of a given computing system. These rules and guidelines, or similar, can be also adopted for evaluating performance portability. In the next section we present our suggestion for extending the SPEC infrastructure for assessing the performance portability of applications and heterogeneous programming models.

SPEC is a three-decade-old consortium formed to develop standardized and realistic benchmark suites for rating and comparing the performance of contemporary computing platforms ranging from a single processor to large-scale supercomputers of thousands of cores.
Three benchmark suites are relevant to the topic of this paper: SPEC ACCEL, SPEC OMP 2012, and SPEChpc 2021, each of which is described below.

The {\bf SPEC ACCEL benchmark} suite was designed to test the performance of computationally intensive parallel applications using three programming models: OpenCL (19 programs), OpenACC (15 programs), and OpenMP 4 target off-loading (15 programs). 

The {\bf SPEC OMP2012 benchmark} suite provides 14 scientific and engineering application codes based on the OpenMP 3.1 standard for measuring the performance of shared-memory parallel machines. The applications were designed in mind to be portable to a variety of CPU architectures and operating systems.

{\bf SPEChpc 2021} provides large-scale scientific applications using the pure MPI standard or hybrid MPI+X, where X can be OpenMP or OpenACC. It contains four suites at different sizes of workload (tiny, small, medium, and large) for evaluating large-scale systems at different sizes, ranging from a single node to hundreds of nodes.

SPEC's methodology is to provide the vendors of computing systems with a simple tool for measuring the performance of their products that will be standardized, objective, and based on strict operating and reporting guidelines. The requirement that the benchmark will be run and reported according to a set of rules makes the results comparable, meaningful, and reproducible. Each benchmark suite is available in source code that has already been ported to various platforms. The source code needs only to be compiled for the target system and then to be tuned for obtaining the best results possible. Each benchmark is comprised of a wide range of representative scientific programs ranging from basic kernels and mini-apps to large weather modeling applications.  

SPEC allows performance tuning at compilation time and at runtime. Performance tuning can be done by using optimal settings of the compiler options or selecting the number of ranks and threads per rank to obtain the best performance.

According to SPEC, two levels of optimization and compilation are allowed:

{\bf Base metrics}. This level enforces strict rules of unaggressive compilation such as using the same flags in the same order for all programs of a given language in a benchmark suite. It demands a common set of optimizations and environment settings to all the programs in a suite, but it allows reordering of arithmetic and floating-point operands. Moreover, at the base level the same compiler must be used for all programs of a given language within a benchmark suite and the same libraries, compiler, and linker options. 

{\bf Peak metrics}. This level is {\bf optional} and allows more flexibility in choosing different compiler options for better performance tuning. At peak level, different compilers may be used for all programs of a given language within the benchmark suite. All flags or options that affect the compilation may be different for each benchmark in the benchmark suite.

In principle, SPEC policy does not allow any modification of the source codes except under specific and restricted circumstances. The SPEC rules are intended to ensure a fair and objective measure of the performance of HPC platforms. For example, SPEC ACCEL allows source code modifications for the peak-level runs of OpenACC and OpenMP benchmarks. Changes to the compiler directives and source code are permitted for portable optimizations to achieve improved scalability. Changes in the algorithm are, however, not permitted. Vendor-specific extensions are allowed if they are portable.

Examples of allowed source code modifications and optimizations are loop reordering, reshaping arrays, and memory distribution. On the other hand, language extensions and adding calls to vendor-specific functions are not allowed.

Furthermore, SPEC allows runtime dynamic optimizations techniques under the control of hardware and software. Such optimizations include improving the instruction cache performance by rearranging the code, value prediction, and reallocation of functional units among hardware threads.

A fundamental principle of SPEC's methodology is given to the requirement of a detailed disclosure of the obtained results and the configuration settings for reproducing benchmark results. Usually, a report of the benchmark results consists of three runs and the median of these runs. It must describe the performance methods that were used and the source-code modifications, if there were any, as well as a general description of each modification applied.

Finally, it is important to note that SPEC encourages using the benchmark suites in academic and research institutions and therefore they are available free of charge for research purposes.

\section{Extending SPEC Repository}

In this section we present the basic concepts and features for upgrading the SPEC infrastructure for rating and comparing the performance portability of applications, benchmarks and models from different perspectives and different application-architecture pair spaces within SPEC repository. 

Before we discuss and specify how performance portability measures can be integrated within the SPEC framework, we have to decide which performance portability metric to apply. Thereafter, we have to decide which performance efficiency approaches we want to use and the performance efficiency types that will be required in order to present the performance portability from different points of view. Finally, we have to recommend which of them will be optional and which ones will be mandatory.

\subsection{Performance portability}

The search for a better performance portability metric is ongoing, and is one of the challenging research areas of the current generation of high-performance heterogeneous computing. The most promising metric proposed to date is the {\funnyPbar} metric \cite{Marowka2}.
The {\funnyPbar} metric is defined as the arithmetic mean of an application's performance efficiency observed across a set of platforms from the same architecture class. Formally, for a given supported set of platforms $S \subseteq H$ from the same architecture class, the performance portability of a case-study application $a$ solving problem $p$ is: 

\begin{align}
{{\funnyPbar}}(a, p, S, H) = 
\begin{cases}
  \frac{\sum_{i \in S}^{} {e_{i}(a, p)}}{|S|}     & \text{if $|S| > 0$} \\ 
  \text{$0$}                                      & \text{otherwise} 
\end{cases}
\end{align}

where $S:= \{i \in H|e_{i}(a, p) > 0\}$ and 
$e_{i}(a, p)$ is the performance efficiency of case-study application $a$ solving problem $p$ on platform $i$.

A comprehensive research study based on dozens of practical studies showed that the {\funnyPbar} metric has the key properties of a good performance portability metric \cite{Marowka2,Marowka3}. These studies show that the {\funnyPbar} metric is objective, comparable, consistent, lossless, easy to use, intuitive, and familiar to users. We recommend adoption of the {\funnyPbar} metric for calculating the performance portability scores of applications tested within the SPEC framework. 
  
We would like to bring to the reader's attention a special added value obtained from the incorporation of the performance portability assessment within the SPEC framework and concerning the set of platforms, $H$, in the definition. From the dozens of studies conducted on the subject of performance portability in recent years, it appears that the average number of platforms on which the studies were based on was four, while the maximum number was 14. Needless to say, the larger the number of platforms, the more accurate the assessment of performance portability. Consolidation of the performance portability assessment within the SPEC framework will increase the number of platforms in $H$ because over time it will include all the platforms that support a given application. Furthermore, the evaluation of the performance portability scores of any given application on any given platform will be done with the same rules and guidelines, and it will be possible to follow the changes of the performance portability over time.

\subsection{Performance efficiency}

Recall the definition of performance portability: 
\begin{quote}
{\it A measurement of an application's performance efficiency for a given problem that can be executed correctly on all platforms in a given set.}
\end{quote}
It follows from the definition that it is based on measuring the performance efficiency of a given application on a specific platform: 

%\smallskip
\begin{quote}
{\bf Definition: Performance Efficiency}

{\it A measurement of an application's achieved performance as a fraction of the baseline performance.}
\end{quote} 
 
when performance is usually measured by runtime or throughput. The baseline performance can be the theoretical or practical peak performance, such as the theoretical peak throughput of a specific GPU or its Roofline peak throughput \cite{Roofline}.

Two performance efficiency approaches have been proposed to date in the scientific literature: application efficiency and architectural efficiency. 

These two approaches present two different perspectives on the relative performance of a given application running on a particular platform and both yield different scores. Each of them examines the performance of a given application in relation to different reference performances. The application efficiency is measured in relation to the performance of the fastest known implementation on that platform, while the architectural performance is measured in relation to the theoretical or practical performance that can possibly be achieved on the given platform.
Now let us define these two approaches formally.

%\smallskip
\begin{quote}
{\bf Definition: application efficiency}

{\it The achieved performance, on a given platform, normalized relative to the best-known performance of an application's implementation on the same platform.} 
\end{quote}

\begin{quote}
{\bf Definition: architectural efficiency}

{\it The application's achieved throughput on a given platform normalized relative to the peak throughput of the given platform.}
\end{quote}

\subsection{Application efficiency approach}
  
SPEC's base metrics and peak metrics are actually the respective equivalents of the achieved performance and peak performance that define the performance efficiency ratio.
Hence, we can define the SPEC efficiency as follows:

\begin{quote}
{\bf Definition: SPEC efficiency}

{\it The ratio of SPEC's base metrics to SPEC's peak metrics.}
\end{quote}

Therefore, the first step that needs to be done in order to extend SPEC for performance portability is to modify the run-rules and the reporting of the results so that the measurements of the peak metrics will not be optional but mandatory, at least for the purpose of calculating performance portability.

Application efficiency is a very popular measure because it is simple and easy to use 
\cite{Daniel,Deakin3,Deakin}. All that is required is to measure the runtime of the application, on the given platform, and then calculate its fraction relative to the runtime of the fastest known portable application on the same platform. The problem is that we can never be sure if we have at hand the fastest implementation. And so, it can happen that immediately after we have published our research, a faster implementation is found which makes the results of our findings outdated.

Furthermore, from the studies that have been done in recent years and which have used this measure, it appears that researchers always chose as the baseline performance the performance of the implementation that showed the best performance from three or four implementations studied in their research and not from those known in the literature \cite{Daniel,Deakin3}. If we add the observation that different studies used different compilers, compiler options, input sizes, and that the source codes are not always available, it is clear that this situation leads to non-uniformity and incoherence of the results and difficulties in reproducing them.

Such situations cannot occur when we restrict ourselves to a rule-based and supervised framework like SPEC. If an implementation with better performance enters the repository, the performance portability calculation of the relevant applications will be automatically updated. Such an automatic update is possible if dynamic web pages are used such as those of a spreadsheet that enables automatic update of the calculation of a given function if one of its variables changes its value.
Such a solution allows for a common performance reference in the repository at any point in time for all applications and benchmark suites. In this way, the database of performance portability reports will remain uniform and consistent while allowing an objective comparison between applications with the possibility of reproducing the various results.

A restricted definition of application efficiency was first introduced in \cite{Marowka3} and was used to calculate the performance portability of portable programming models. The definition was formulated after a survey based on hundreds of case studies which showed that most researchers use this measure in practice. This measure reflects how far the performance of a given portable application is from the peak practical performance possible, or in other words, the cost in performance that a portable application sacrifices to be portable.

Therefore, in order to integrate this measure into the SPEC framework, the best performance of a low-level, unportable, and optimized implementation that appears in the SPEC repository needs to be selected as the baseline performance. In addition, it is required that if a faster unportable and optimized implementation will appear in the future in SPEC, an automatic update of the performance portability scores of all relevant applications in the SPEC's repository will be updated accordingly.

We define three types of performance efficiency according to a reference application whose performance is used as the baseline performance. In each of the efficiency types, the reference application has a different level of abstraction, so its performance is directly derived from its ability to utilize the hardware resources of the platform effectively.  
The following application efficiency types are described in increasing order of the peak achievable performance of the reference application.

\begin{quote}
{\bf Definition: application efficiency-Type 0

     (SPEC efficiency)}

{\it The achieved SPEC's base metrics of a given portable application-platform pair normalized relative to the SPEC's peak metrics on the same application-platform pair.}
\end{quote}

All SPEC's run-rules and guidelines apply for measuring this type of performance efficiency. It yields high values since the optimization level of the SPEC's peak metrics is usually restricted to choose different compiler options for better performance tuning or by making changes to the compiler directives.

In the next section we will demonstrate, using SPEC efficiency, how to calculate the performance portability of applications and benchmarks from data taken from the current SPEC repository.

\bigskip
\bigskip

\begin{quote}
{\bf Definition: application efficiency-Type 1}

{\it The achieved performance of a given portable application-platform pair, normalized relative to the best-known performance of any portable application on the same platform in the SPEC repository.}
\end{quote}

Here the baseline performance is the performance of any implementation of the application that uses another performance portability framework that achieved the best performance on the same platform within SPEC repository.  For example, the performance of an OpenACC implementation, on an NVIDIA V100 GPU, normalized relative to the performance of a Kokkos implementation that outperforms the OpenACC implementation on NVIDIA V100.

This type of application efficiency expands the space of the application's implementations from which the best baseline performance can be chosen. This space includes all the implementations of the application in any performance portability framework on the same platform within SPEC repository.

\begin{quote}
{\bf Definition: application efficiency-Type 2}

{\it The achieved performance of a given portable application-platform pair, normalized relative to the best-known performance of any unportable application on the same platform in the SPEC repository.}
\end{quote}

This type of application efficiency expands the space of the application's implementations even further. Here the baseline performance can be the performance of any application's implementation on the same platform, and not necessarily a portable one. For example, the performance of an OpenACC implementation on an NVIDIA V100 normalized against a CUDA implementation that outperforms the implementation of OpenACC on NVIDIA V100.

\subsection{Architectural efficiency approach}
  
Architectural efficiency measures the extent to which the application utilizes the resources of the platform on which it is implemented in relation to two reference levels of performance: one is the peak theoretically possible performance level, that is, an unattainable upper-bound performance level, and the other is the practical peak performance level, that is, a performance level which can be achieved through the optimization of all platform resources. Therefore, we distinguish between two types of performance efficiency measures accordingly to the peak performance reference used: theoretical peak throughput or practical peak throughput.

\begin{quote}
{\bf Definition: architectural efficiency-Type 0}

{\it The achieved throughput of a given portable application-platform pair, normalized relative to the peak theoretical throughput of the given platform.}
\end{quote}
  
Architectural efficiency is relatively simple to measure. All that needs to be done is to measure the throughput, in GFLOP/s or GB/s, of the application using a profiling tool and then calculate its fraction relative to the theoretical performance published by the vendor.
Practitioners do not like this measure because its results yield a theoretical score. Therefore, they prefer more practical measure such as using the Roofline model.

\onecolumn

\begin{table}
\caption{Summary of the different types of the performance efficiency approaches.}
\begin{center}
\footnotesize
%\resizebox{\columnwidth}{0.06\textheight}{% % !
%\resizebox{\columnwidth}{!}{% 

%{
%\setlength\arrayrulewidth{0.5pt}

\begin{tabular}{|l|l|l|l|l|l|l|}
\hline  
\multicolumn{1}{c|}{}
&\multicolumn{6}{c}{Performance Efficiency Approach}\\
\hline
\multicolumn{1}{c|}{}
&\multicolumn{3}{c|}{Application Efficiency}
&\multicolumn{3}{c}{Architectural Efficiency} \\ %\cline{1-8}
\hline
\multicolumn{1}{c|}{}
&\multicolumn{3}{c|}{Relative Baseline Application}
&\multicolumn{3}{c}{Relative Baseline Application} \\ %\cline{1-8}
\hline
\multicolumn{1}{c|}{Type}
&\multicolumn{1}{c|}{Application}
&\multicolumn{1}{c|}{Platform} 
&\multicolumn{1}{c|}{Performance} 
&\multicolumn{1}{c|}{Application}
&\multicolumn{1}{c|}{Platform} 
&\multicolumn{1}{c}{Performance} \\
\hline
\hline
\multicolumn{1}{c|}{0}
&\multicolumn{1}{c|}{same}
&\multicolumn{1}{c|}{same} 
&\multicolumn{1}{c|}{peak metrics} 
&\multicolumn{1}{c|}{same}
&\multicolumn{1}{c|}{same} 
&\multicolumn{1}{c}{peak theoretical} \\
\hline
\multicolumn{1}{c|}{1}
&\multicolumn{1}{c|}{any portable}
&\multicolumn{1}{c|}{same} 
&\multicolumn{1}{c|}{best-known} 
&\multicolumn{1}{c|}{same}
&\multicolumn{1}{c|}{same} 
&\multicolumn{1}{c}{peak Roofline} \\
\hline
\multicolumn{1}{c|}{2}
&\multicolumn{1}{c|}{any unportable}
&\multicolumn{1}{c|}{same} 
&\multicolumn{1}{c|}{best-known} 
&\multicolumn{1}{c|}{-}
&\multicolumn{1}{c|}{-} 
&\multicolumn{1}{c}{-} \\
\hline
\end{tabular}
%}%}
\label{table120}
\end{center}
\end{table}

\begin{table}
\caption{The list of SPEC OMP2012 applications.}
\begin{center}
\footnotesize
%\resizebox{\columnwidth}{0.06\textheight}{% % !
%\resizebox{\columnwidth}{!}{% 

%{
%\setlength\arrayrulewidth{0.5pt}

\begin{tabular}{|l|l|l|}
\hline  
\multicolumn{1}{c|}{Benchmark}
&\multicolumn{1}{c|}{Language}
&\multicolumn{1}{c}{Application domain} \\
\hline
\multicolumn{1}{c|}{350.md}
&\multicolumn{1}{c|}{Fortran}
&\multicolumn{1}{c}{Molecular Dynamics} \\ %\cline{1-8}
\hline
\multicolumn{1}{c|}{351.bwaves}
&\multicolumn{1}{c|}{Fortran}
&\multicolumn{1}{c}{Fluid Dynamics} \\ %\cline{1-8}
\hline
\multicolumn{1}{c|}{352.nab}
&\multicolumn{1}{c|}{C}
&\multicolumn{1}{c}{Molecular Modeling} \\ %\cline{1-8}
\hline
\multicolumn{1}{c|}{357.bt331}
&\multicolumn{1}{c|}{Fortran}
&\multicolumn{1}{c}{Fluid Dynamics} \\ %\cline{1-8}
\hline
\multicolumn{1}{c|}{358.botsalgn}
&\multicolumn{1}{c|}{C}
&\multicolumn{1}{c}{Protein Alignment} \\ %\cline{1-8}
\hline
\multicolumn{1}{c|}{359.botsspar}
&\multicolumn{1}{c|}{C}
&\multicolumn{1}{c}{Sparse LU} \\ %\cline{1-8}
\hline
\multicolumn{1}{c|}{360.ilbdc}
&\multicolumn{1}{c|}{Fortran}
&\multicolumn{1}{c}{Lattic Boltzmann} \\ %\cline{1-8}
\hline
\multicolumn{1}{c|}{362.fma3d}
&\multicolumn{1}{c|}{Fortran}
&\multicolumn{1}{c}{Mechanical Simulation} \\ %\cline{1-8}
\hline
\multicolumn{1}{c|}{363.swim}
&\multicolumn{1}{c|}{Fortran}
&\multicolumn{1}{c}{Weather Prediction} \\ %\cline{1-8}
\hline
\multicolumn{1}{c|}{367.imagick}
&\multicolumn{1}{c|}{C}
&\multicolumn{1}{c}{Image Processing} \\ %\cline{1-8}
\hline
\multicolumn{1}{c|}{370.mgrid3311}
&\multicolumn{1}{c|}{Fortran}
&\multicolumn{1}{c}{Fluid Dynamics} \\ %\cline{1-8}
\hline
\multicolumn{1}{c|}{371.applu331}
&\multicolumn{1}{c|}{Fortran}
&\multicolumn{1}{c}{Fluid Dynamics} \\ %\cline{1-8}
\hline
\multicolumn{1}{c|}{372.smithwa}
&\multicolumn{1}{c|}{C}
&\multicolumn{1}{c}{Pattern Matching} \\ %\cline{1-8}
\hline
\multicolumn{1}{c|}{376.kdtree}
&\multicolumn{1}{c|}{C++}
&\multicolumn{1}{c}{Sorting and Searching} \\ %\cline{1-8}
\hline
\end{tabular}
%}
\label{table1}
\end{center}
\end{table}

\begin{table}
\caption{The list of platforms used for the case study and their configuration.}
\begin{center}
\footnotesize
%\resizebox{\columnwidth}{0.06\textheight}{% % !
%\resizebox{\columnwidth}{!}{% 

%{
%\setlength\arrayrulewidth{0.5pt}

\begin{tabular}{|l|l|l|}
\hline  
\multicolumn{1}{c|}{Platform No.}
&\multicolumn{1}{c|}{Platform}
&\multicolumn{1}{c}{Configuration} \\
\hline
\multicolumn{1}{c|}{1}
&\multicolumn{1}{c|}{Intel Xeon E5-2670}
&\multicolumn{1}{c}{16 cores, 2 chips, 8 cores/chip} \\ %\cline{1-8}
\hline
\multicolumn{1}{c|}{2}
&\multicolumn{1}{c|}{Intel Xeon E5-2697 v2}
&\multicolumn{1}{c}{24 cores, 2 chips, 12 cores/chip} \\ %\cline{1-8}
\hline
\multicolumn{1}{c|}{3}
&\multicolumn{1}{c|}{Intel Xeon E7-8890 v3}
&\multicolumn{1}{c}{72 cores, 4 chips, 18 cores/chip} \\ %\cline{1-8}
\hline
\multicolumn{1}{c|}{4}
&\multicolumn{1}{c|}{Intel Xeon E7-8890 v3}
&\multicolumn{1}{c}{288 cores, 16 chips, 18 cores/chip} \\ %\cline{1-8}
\hline
\multicolumn{1}{c|}{5}
&\multicolumn{1}{c|}{Intel Xeon Phi 7210}
&\multicolumn{1}{c}{64 cores, 1 chip, 64 cores/chip} \\ %\cline{1-8}
\hline
\multicolumn{1}{c|}{6}
&\multicolumn{1}{c|}{Intel Xeon Gold 6154}
&\multicolumn{1}{c}{576 cores, 32 chips, 18 cores/chip} \\ %\cline{1-8}
\hline
\multicolumn{1}{c|}{7}
&\multicolumn{1}{c|}{Intel Xeon Platinum 8260L}
&\multicolumn{1}{c}{48 cores, 2 chips, 24 cores/chip} \\ %\cline{1-8}
\hline
\multicolumn{1}{c|}{8}
&\multicolumn{1}{c|}{Intel Xeon Platinum 9242}
&\multicolumn{1}{c}{96 cores, 2 chips, 48 cores/chip} \\ %\cline{1-8}
\hline
\multicolumn{1}{c|}{9}
&\multicolumn{1}{c|}{AMD EPYC 9654}
&\multicolumn{1}{c}{192 cores, 2 chips, 96 cores/chip} \\ %\cline{1-8}
\hline
\multicolumn{1}{c|}{10}
&\multicolumn{1}{c|}{SPARC T7-4}
&\multicolumn{1}{c}{128 cores, 4 chips, 32 cores/chip} \\ %\cline{1-8}
\hline
\end{tabular}
%}
\label{table12}
\end{center}
\end{table} 

\twocolumn

\begin{quote}
{\bf Definition: architectural efficiency-Type 1}

{\it The achieved throughput of a given portable application-platform pair, normalized relative to the peak Roofline throughput of the given platform.}
\end{quote}  

The Roofline model is a visualization tool that shows the type of peak throughput that might be expected for an application with a given arithmetic intensity. The Roofline graph is a line whose slope is associated with the peak memory bandwidth throughput (GB/s), and then a flat part that is associated with peak flop throughput (GFLOP/s).

Unfortunately, it is a time-consuming and challenging task to estimate the platform features needed for a Roofline analysis \cite{Bertoni}. Moreover, due to the lack of standardization of the profile tools and the progressively optimized micro-benchmarks used for generating Roofline graphs, multiple graphs tend to be created with different properties for the same platform \cite{Marowka2}. 
 
This problem can be solved by very rigorous rules and guidelines that will dictate how Roofline graphs should be created. These rules will determine in detail which profiling tools and progressively optimized micro-benchmarks to use for generating Roofline graphs and which platform features are needed for a Roofline analysis. There are tools on the market that can greatly facilitate the process of creating a Roofline graph, for example Intel Vtune \cite{Vtune} or Empirical Roofline Tool (ERT) \cite{ERT}. At the end of the process, SPEC committee members will approve which Roofline graph to use for measuring the Roofline efficiency for all SPEC applications.

{\bf Bottom line}. The performance efficiency types presented in this section enlighten different and complementary perspectives of an application's performance portability. At the same time, multiple types can sometimes be confusing rather than helpful. It is certainly possible to choose fewer performance efficiency types or to decide that some of them will be mandatory and others optional. We will leave this decision to the SPEC committee as part of the drafting of the final document. Table II presents a concise summary and comparison of the different types of the performance efficiency approaches.

In the next section we show and demonstrate how to calculate the performance portability of applications and benchmark suites using the SPEC efficiency  and {\funnyPbar} metric.

\section{Examples based on SPEC Repository}

In this section we present examples based on the performance of applications of SPEC OMP 2012 benchmark that appear within the current SPEC repository. We calculate the performance portability score of three applications and of the whole benchmark itself. 
We used SPEC's performance efficiency and the {\funnyPbar} metric for calculating the performance portability scores.

Since the current SPEC repository is performance oriented and not performance portability, we were forced to present fewer examples than we would like. For example, the performance reporting of the most of the platforms in the current SPEC repository does not include the {\it SPEC's peak metrics} since the reporting of this performance score is optional. Therefore, we were unable to present examples of additional programming models, such as OpenACC, on state-of-the-art platforms. 
%including all the performance efficiency types. 
Furthermore, the performance portability scores presented in this paper were calculated only for the SPEC Efficiency since the current SPEC repository lacks the data needed to calculate the performance portability based on all the performance efficiency approaches and their types.
However, the purpose of the examples is primarily to demonstrate the ideas presented in this paper.

\begin{table}
\caption{Performance Portability of the Molecular Dynamics application.}
\begin{center}
\footnotesize
%\resizebox{\columnwidth}{0.06\textheight}{% % !
%\resizebox{\columnwidth}{!}{% 

%{
%\setlength\arrayrulewidth{0.5pt}

\begin{tabular}{|l|l|l|l|l|l|}
\hline  
\multicolumn{6}{c}{350.md Molecular Dynamics} \\
\hline  
\multicolumn{1}{c|}{Platform No.}
&\multicolumn{2}{c|}{Base}
&\multicolumn{2}{c|}{Peak} 
&\multicolumn{1}{c}{Efficiency}\\
\hline
\multicolumn{1}{c|}{}
&\multicolumn{1}{c|}{threads}
&\multicolumn{1}{c|}{seconds} 
&\multicolumn{1}{c|}{thread}
&\multicolumn{1}{c|}{seconds}
&\multicolumn{1}{c}{\%} \\
\hline
\multicolumn{1}{c|}{1}
&\multicolumn{1}{c|}{32}
&\multicolumn{1}{c|}{975} 
&\multicolumn{1}{c|}{32}
&\multicolumn{1}{c|}{803}
&\multicolumn{1}{c}{82} \\
\hline
\multicolumn{1}{c|}{2}
&\multicolumn{1}{c|}{48}
&\multicolumn{1}{c|}{585} 
&\multicolumn{1}{c|}{48}
&\multicolumn{1}{c|}{483}
&\multicolumn{1}{c}{83} \\
\hline
\multicolumn{1}{c|}{3}
&\multicolumn{1}{c|}{144}
&\multicolumn{1}{c|}{197} 
&\multicolumn{1}{c|}{144}
&\multicolumn{1}{c|}{161}
&\multicolumn{1}{c}{82} \\
\hline
\multicolumn{1}{c|}{4}
&\multicolumn{1}{c|}{576}
&\multicolumn{1}{c|}{59.5} 
&\multicolumn{1}{c|}{576}
&\multicolumn{1}{c|}{38.6}
&\multicolumn{1}{c}{65} \\
\hline
\multicolumn{1}{c|}{5}
&\multicolumn{1}{c|}{256}
&\multicolumn{1}{c|}{537} 
&\multicolumn{1}{c|}{256}
&\multicolumn{1}{c|}{434}
&\multicolumn{1}{c}{81} \\
\hline
\multicolumn{1}{c|}{6}
&\multicolumn{1}{c|}{513}
&\multicolumn{1}{c|}{5.6} 
&\multicolumn{1}{c|}{576}
&\multicolumn{1}{c|}{5.33}
&\multicolumn{1}{c}{95} \\
\hline
\multicolumn{1}{c|}{7}
&\multicolumn{1}{c|}{96}
&\multicolumn{1}{c|}{33.4} 
&\multicolumn{1}{c|}{96}
&\multicolumn{1}{c|}{33.3}
&\multicolumn{1}{c}{99} \\
\hline
\multicolumn{1}{c|}{8}
&\multicolumn{1}{c|}{192}
&\multicolumn{1}{c|}{16.9} 
&\multicolumn{1}{c|}{192}
&\multicolumn{1}{c|}{16.8}
&\multicolumn{1}{c}{99} \\
\hline
\multicolumn{1}{c|}{9}
&\multicolumn{1}{c|}{384}
&\multicolumn{1}{c|}{31.3} 
&\multicolumn{1}{c|}{192}
&\multicolumn{1}{c|}{30.5}
&\multicolumn{1}{c}{97} \\
\hline
\multicolumn{1}{c|}{10}
&\multicolumn{1}{c|}{256}
&\multicolumn{1}{c|}{153} 
&\multicolumn{1}{c|}{768}
&\multicolumn{1}{c|}{111}
&\multicolumn{1}{c}{72} \\
\hline
\multicolumn{6}{c}{}\\
\multicolumn{6}{c}{{\funnyPbar} = 85.5\%}\\
\hline
\end{tabular}
%}
\label{table2}
\end{center}
\end{table} 

\begin{table}
\caption{Performance Portability of the Protein Alignment application.}
\begin{center}
\footnotesize
%\resizebox{\columnwidth}{0.06\textheight}{% % !
%\resizebox{\columnwidth}{!}{% 

%{
%\setlength\arrayrulewidth{0.5pt}

\begin{tabular}{|l|l|l|l|l|l|}
\hline  
\multicolumn{6}{c}{358.botsalgn Protein Alignment} \\
\hline  
\multicolumn{1}{c|}{Platform No.}
&\multicolumn{2}{c|}{Base}
&\multicolumn{2}{c|}{Peak} 
&\multicolumn{1}{c}{Efficiency}\\
\hline
\multicolumn{1}{c|}{}
&\multicolumn{1}{c|}{threads}
&\multicolumn{1}{c|}{seconds} 
&\multicolumn{1}{c|}{thread}
&\multicolumn{1}{c|}{seconds}
&\multicolumn{1}{c}{\%} \\
\hline
\multicolumn{1}{c|}{1}
&\multicolumn{1}{c|}{32}
&\multicolumn{1}{c|}{1276} 
&\multicolumn{1}{c|}{32}
&\multicolumn{1}{c|}{1235}
&\multicolumn{1}{c}{97} \\
\hline
\multicolumn{1}{c|}{2}
&\multicolumn{1}{c|}{48}
&\multicolumn{1}{c|}{808} 
&\multicolumn{1}{c|}{48}
&\multicolumn{1}{c|}{779}
&\multicolumn{1}{c}{96} \\
\hline
\multicolumn{1}{c|}{3}
&\multicolumn{1}{c|}{144}
&\multicolumn{1}{c|}{287} 
&\multicolumn{1}{c|}{144}
&\multicolumn{1}{c|}{280}
&\multicolumn{1}{c}{98} \\
\hline
\multicolumn{1}{c|}{4}
&\multicolumn{1}{c|}{576}
&\multicolumn{1}{c|}{74.6} 
&\multicolumn{1}{c|}{576}
&\multicolumn{1}{c|}{74.5}
&\multicolumn{1}{c}{99} \\
\hline
\multicolumn{1}{c|}{5}
&\multicolumn{1}{c|}{256}
&\multicolumn{1}{c|}{1133} 
&\multicolumn{1}{c|}{256}
&\multicolumn{1}{c|}{1136}
&\multicolumn{1}{c}{100} \\
\hline
\multicolumn{1}{c|}{6}
&\multicolumn{1}{c|}{513}
&\multicolumn{1}{c|}{29.5} 
&\multicolumn{1}{c|}{576}
&\multicolumn{1}{c|}{26.7}
&\multicolumn{1}{c}{90} \\
\hline
\multicolumn{1}{c|}{7}
&\multicolumn{1}{c|}{96}
&\multicolumn{1}{c|}{304} 
&\multicolumn{1}{c|}{96}
&\multicolumn{1}{c|}{286}
&\multicolumn{1}{c}{94} \\
\hline
\multicolumn{1}{c|}{8}
&\multicolumn{1}{c|}{192}
&\multicolumn{1}{c|}{141} 
&\multicolumn{1}{c|}{192}
&\multicolumn{1}{c|}{136}
&\multicolumn{1}{c}{96} \\
\hline
\multicolumn{1}{c|}{9}
&\multicolumn{1}{c|}{384}
&\multicolumn{1}{c|}{49.8} 
&\multicolumn{1}{c|}{384}
&\multicolumn{1}{c|}{49.8}
&\multicolumn{1}{c}{100} \\
\hline
\multicolumn{1}{c|}{10}
&\multicolumn{1}{c|}{256}
&\multicolumn{1}{c|}{166} 
&\multicolumn{1}{c|}{256}
&\multicolumn{1}{c|}{165}
&\multicolumn{1}{c}{99} \\
\hline
\multicolumn{6}{c}{}\\
\multicolumn{6}{c}{{\funnyPbar} = 96.9\%}\\
\hline
\end{tabular}
%}
\label{table3}
\end{center}
\end{table} 

\begin{table}
\caption{Performance Portability of the Weather Prediction  application.}
\begin{center}
\footnotesize
%\resizebox{\columnwidth}{0.06\textheight}{% % !
%\resizebox{\columnwidth}{!}{% 

%{
%\setlength\arrayrulewidth{0.5pt}

\begin{tabular}{|l|l|l|l|l|l|}
\hline  
\multicolumn{6}{c}{363.swim   Weather Prediction } \\
\hline  
\multicolumn{1}{c|}{Platform No.}
&\multicolumn{2}{c|}{Base}
&\multicolumn{2}{c|}{Peak} 
&\multicolumn{1}{c}{Efficiency}\\
\hline
\multicolumn{1}{c|}{}
&\multicolumn{1}{c|}{threads}
&\multicolumn{1}{c|}{seconds} 
&\multicolumn{1}{c|}{thread}
&\multicolumn{1}{c|}{seconds}
&\multicolumn{1}{c}{\%} \\
\hline
\multicolumn{1}{c|}{1}
&\multicolumn{1}{c|}{32}
&\multicolumn{1}{c|}{855} 
&\multicolumn{1}{c|}{16}
&\multicolumn{1}{c|}{771}
&\multicolumn{1}{c}{90} \\
\hline
\multicolumn{1}{c|}{2}
&\multicolumn{1}{c|}{48}
&\multicolumn{1}{c|}{667} 
&\multicolumn{1}{c|}{24}
&\multicolumn{1}{c|}{608}
&\multicolumn{1}{c}{91} \\
\hline
\multicolumn{1}{c|}{3}
&\multicolumn{1}{c|}{144}
&\multicolumn{1}{c|}{219} 
&\multicolumn{1}{c|}{72}
&\multicolumn{1}{c|}{212}
&\multicolumn{1}{c}{96} \\
\hline
\multicolumn{1}{c|}{4}
&\multicolumn{1}{c|}{576}
&\multicolumn{1}{c|}{79.3} 
&\multicolumn{1}{c|}{288}
&\multicolumn{1}{c|}{77.8}
&\multicolumn{1}{c}{97} \\
\hline
\multicolumn{1}{c|}{5}
&\multicolumn{1}{c|}{256}
&\multicolumn{1}{c|}{233} 
&\multicolumn{1}{c|}{256}
&\multicolumn{1}{c|}{220}
&\multicolumn{1}{c}{94} \\
\hline
\multicolumn{1}{c|}{6}
&\multicolumn{1}{c|}{513}
&\multicolumn{1}{c|}{28.4} 
&\multicolumn{1}{c|}{567}
&\multicolumn{1}{c|}{26.5}
&\multicolumn{1}{c}{90} \\
\hline
\multicolumn{1}{c|}{7}
&\multicolumn{1}{c|}{96}
&\multicolumn{1}{c|}{310} 
&\multicolumn{1}{c|}{48}
&\multicolumn{1}{c|}{298}
&\multicolumn{1}{c}{96} \\
\hline
\multicolumn{1}{c|}{8}
&\multicolumn{1}{c|}{192}
&\multicolumn{1}{c|}{153} 
&\multicolumn{1}{c|}{96}
&\multicolumn{1}{c|}{145}
&\multicolumn{1}{c}{94} \\
\hline
\multicolumn{1}{c|}{9}
&\multicolumn{1}{c|}{384}
&\multicolumn{1}{c|}{87.9} 
&\multicolumn{1}{c|}{192}
&\multicolumn{1}{c|}{82.9}
&\multicolumn{1}{c}{94} \\
\hline
\multicolumn{1}{c|}{10}
&\multicolumn{1}{c|}{256}
&\multicolumn{1}{c|}{146} 
&\multicolumn{1}{c|}{128}
&\multicolumn{1}{c|}{140}
&\multicolumn{1}{c}{95} \\
\hline
\multicolumn{6}{c}{}\\
\multicolumn{6}{c}{{\funnyPbar} = 93.7\%}\\
\hline
\end{tabular}
%}
\label{table4}
\end{center}
\end{table}

\begin{table}
\caption{Performance Portability of SPEC OMP2012 suite.}
\begin{center}
\footnotesize
%\resizebox{\columnwidth}{0.06\textheight}{% % !
%\resizebox{\columnwidth}{!}{% 

%{
%\setlength\arrayrulewidth{0.5pt}

\begin{tabular}{|l|l|}
\hline  
\multicolumn{2}{c}{SPEC OMP 2012} \\
\hline  
\multicolumn{1}{c|}{Platform No.}
&\multicolumn{1}{c}{Efficiency \%}\\
\hline
\multicolumn{1}{c|}{1}
&\multicolumn{1}{c}{94} \\
\hline
\multicolumn{1}{c|}{2}
&\multicolumn{1}{c}{91} \\
\hline
\multicolumn{1}{c|}{3}
&\multicolumn{1}{c}{94} \\
\hline
\multicolumn{1}{c|}{4}
&\multicolumn{1}{c}{86} \\
\hline
\multicolumn{1}{c|}{5}
&\multicolumn{1}{c}{98} \\
\hline
\multicolumn{1}{c|}{6}
&\multicolumn{1}{c}{95} \\
\hline
\multicolumn{1}{c|}{7}
&\multicolumn{1}{c}{82} \\
\hline
\multicolumn{1}{c|}{8}
&\multicolumn{1}{c}{84} \\
\hline
\multicolumn{1}{c|}{9}
&\multicolumn{1}{c}{96} \\
\hline
\multicolumn{1}{c|}{10}
&\multicolumn{1}{c}{94} \\
\hline
\multicolumn{2}{c}{}\\
\multicolumn{2}{c}{{\funnyPbar} = 91.4\%}\\
\hline
\end{tabular}
%}
\label{table5}
\end{center}
\end{table} 

Table III shows the 14 applications of the SPEC OMP 2012 benchmark suite written using OpenMP 3.1 with a short description of the domain of each one of the applications. Table IV shows the 10 platforms that were used for our examples  and their configurations. It can be observed that all the platforms are SMP machines with 16 cores and up to 576 cores. Tables V, VI, and VII show the SPEC performance efficiencies measured for molecular dynamics, protein alignment, and weather prediction applications, respectively. The performance portability scores that were obtained are 85.5\%, 96.9\%, and 93.7\%, respectively, which are considered high scores but quite expected because the reference performance of the SPEC efficiency was not achieved after aggressive optimizations. Table VIII shows the performance portability score, 91.4\%, of the whole SPEC OMP 2012 suite on the given platforms.

\section{Conclusions}

The extensive collection of independent studies done in recent years to study the performance portability of applications is not based on common rules and guidelines. As a result, there is great difficulty in comparing the findings of the various studies in order to reach informed conclusions and insights that will allow software and hardware architects to improve the performance portability and productivity of scientific applications in the future in light of the constant acceleration in technological innovations and the design of heterogeneous systems.
 
In this paper we have presented a proposal for building an appropriate repository for performance portability within an existing SPEC framework. 
Such a repository will be standardized, objective, and based on strict operating and reporting guidelines. Such guidelines will ensure a fair, comparable and meaningful measure of the performance portability while the requirement for a detailed disclosure of the obtained results and the configuration settings will ensure the reproducibility of the reported results.

We also demonstrated how to calculate the performance portability of applications and of an entire benchmark suite that are currently available in SPEC repository. 
In our future work, we plan to develop a series of benchmarks in order to present an effective comparison of different performance efficiency approaches to calculate the performance portability of applications, benchmarks and models based on the definitions presented in this paper.

\end{document}